# Effects of Elliptically Polarizing Undulator on Beam Dynamics in PLS-II

**S. Shin, D-E. Kim and J-Y. Huang**

***Pohang Accelerator Laboratory, POSTECH, Pohang, 790-784, KOREA***

**S. Chunjarean**

***Chiang Mai University, 239 Huay Kaew Road, Muang District, Chiang Mai, Thailand, 50200***

The non-linear effects caused by the intrinsic field's transverse roll-off in an Elliptically Polarizing Undulator with 72-mm period and maximum peak field of 0.72 T (EPU72) on the dynamics aperture of the Pohang Light Source II (PLS-II) were investigated. A kick map and Frequency map analysis both demonstrated that EPU72 will not reduce the lifetime or cause injection problems in PLS-II.

PACS number: 07-85-Qe

Keywords: Non-linear effect, Kick map, Frequency map analysis

Email: huang@postech.ac.kr

Fax: +82-54-279-1468

## I. INTRODUCTION

The 2.5 GeV storage ring at the Pohang Light Source (PLS) [1] had been operational since 1995, although initially at 2.0 GeV. During this time, insertion devices, such as wigglers and undulator magnets, have added to produce high flux photon beams. The 30 photon beam lines that have been installed were in operation, and ~ 3000 experiments had been conducted since 1995 by researchers from Korea and other countries.

To meet increasing user requirements, the Pohang Accelerator Laboratory (PAL) started the PLS-II project in January 2009 [2, 3] to upgrade (Table 1) the PLS facilities and overcome their limitations. The PLS-II project was completed in March 21, 2012: beam energy was increased from 2.5 GeV to 3.0 GeV, and the stored beam current is being increased from 200 mA to 400 mA. The emittance was decreased from 18.9 nm at 2.5 GeV to 5.8 nm at 3 GeV while the existing storage ring tunnel structure remained. Top-up mode operation [4] was used to stabilize the synchrotron radiation flux.

Insertion devices can be inserted in twenty straight sections. Planar in-vacuum undulator, out-vacuum undulators, a high field multi-pole wiggler and elliptically polarizing undulators have been installed as insertion devices in PLS-II. The fundamental wavelength radiated from an insertion device is

$$\lambda_1(\theta) = \frac{\lambda_w}{2\gamma^2}(1 + \frac{K^2}{2} + \gamma^2\theta^2), \tag{1}$$

where $K$ [dimensionless] is the field strength parameter, $\lambda_w$ [m] is the length of one period of the magnetic field and $\theta$ [rad] is radiation observation angle. In general, an in-vacuum undulator is used as the X-ray source, and an out-vacuum undulator such as an elliptically polarizing undulator (EPU) is used as the ultra-violet (UV) source. However, the available brightness from the undulator is usually $10^5$ times higher than the brightness from a bending magnet. To control the polarization of the photon beam, PLS-II has been fitted with an EPU that can generate bright photon beams with variable linear, elliptical or circular polarization characteristics. As result, many experiments in many fields, including spin-dependent electronic structures of semimetals, magnetic origin of transition metal-oxides, magnetic properties of diluted magnetic semiconductors have been conducted on the EPU beam-line in PLS-II.

One of the major problems with the beam dynamics of third generation synchrotron light source storage rings is the difficulty in evaluating and controlling both linear and nonlinear effects introduced by insertion devices. The linear effect is mainly associated with distortion of beam optics, and can be controlled by adjusting two or three quadrupoles adjacent to the insertion device [5]. The nonlinear effect is a consequence of the nonlinear magnetic fields of the insertion device; this effect is difficult to evaluate and in general it is not controllable.

The vertical field distribution along the EPU axis varies along the horizontal axis; this trend is called transverse field roll-off (TFR). EPU in various insertion devices exhibit a fast TFR of their magnetic fields. Combined with the undulating motion of the beam, the intrinsic field's TFR causes dynamic multi-poles that affect both the nonlinear focusing and the linear transverse dynamics. Therefore, the effects of EPU on the motion of a stored beam must be quantified. We present in this paper the effects of the EPU on beam dynamics in PLS-II. Magnetic field modeling and kick map analysis [6] are introduced in Section 2 and 3, respectively. The result of frequency map analysis will be described in Section 4, and conclusions are given in Section 5.

## II. MAGNETIC FIELD OF EPU

An EPU with 7.2-cm period and 0.72-T maximum peak field (EPU72) has been designed (Table 2) to generate polarized undulator radiation ranging from 78 eV to about a few kiloelectronvolts. The EPU72 undulator is an APPLE-II type consisting of four standard Halbach-type magnet arrays, two above the mid-plane and two below it. This type of EPU has the advantage of open mid-plane for synchrotron extraction and installation of an antechamber. The shifting of the diagonal arrays relative to the fixed arrays can produce horizontal, circular and vertical modes of polarization of the magnetic field. The magnetic field on the axis of the EPU can be expressed as

$$\begin{aligned} B_y(0,0,z) &= B_{y0}\cos\left(\frac{\phi}{2}\right)\sin\left(\frac{2\pi}{\lambda_p}z+\frac{\phi}{2}\right) \\ B_x(0,0,z) &= -B_{x0}\sin\left(\frac{\phi}{2}\right)\cos\left(\frac{2\pi}{\lambda_p}z+\frac{\phi}{2}\right) \end{aligned} \quad (2)$$

with undulator phase $\phi = 2\pi\left(\Delta z/\lambda_p\right)$, where $\Delta z$ is the longitudinal translation of the diagonal arrays, and $B_{x0}$ and $B_{y0}$ are effective peak fields for horizontal and vertical planes, respectively. The effective peak field at the minimum gap of 18 mm is affected by polarization mode (Table 3). At $\Delta z = 0$, only the vertical field exists, so the electrons generate horizontally-polarized radiation. By moving the two diagonal rows 20.87 mm which corresponds to the circular mode, both $B_x$ and $B_y$ on the axis are produced (Fig. 1). To achieve vertical polarization, the two diagonal rows should be translated by half of the undulator period of 36 mm which corresponds to a phase of 180°. In vertical polarization mode, only $B_x$ with amplitude 0.563 T exists. Because the left and right arrays should be translated, a small (~ 1.0 mm) horizontal clearance centered at x = 0 is left between the left and right arrays. Due to this horizontal gap, the vertical field at zero phase as function of x has a dip at x = 0 (Fig. 2).

An electron that moves off-axis through such a device will experience different field strength than experienced by an electron that moves on the axis. The result is a focusing effect imparts an angular kick to the off-centered electron beam. These effects can reduce the dynamic aperture. To evaluate this effect, the magnetic field of the EPU undulator is modelled using RADIA [7]. Then the calculated fields are used in a kick map and tracking code to study the dynamic aperture of the storage ring.

## III. KICK MAP ANALYSIS

The interaction of the EPU with the stored electron beam can be classified as three perturbations. Often users adjust the spectrum from undulators by changing undulator gaps or row phase in the EPU. To avoid disrupting other users, the orbit must be kept constant during these field changes. Usually two steering magnets are used to correct the first and second field integrals. The EPU imparts tune shift caused by the quadrupole term, and coupling error caused by a shift-dependent skew quadrupole term. These effects are corrected by adjusting quadrupoles and skew magnet near the EPU. An EPU intrinsically exhibits a fast TFR of its magnetic field. Combined with the undulating motion of the beam, field roll-off induces formation of dynamic multi-poles that affect the linear focusing and the nonlinear transverse dynamics.

The EPU field generates both focusing and defocusing effects on the electron beam. These effects on the electrons vary according to their transverse entering position, so the field integral becomes non-zero. The combination of non-zero field integral and electron wiggling provides an angular deflection that strongly depends on trajectory offset. As result, the TFR is introduced. Kick map modelling of the EPU is used to calculate the deflecting angle for the particle tracking through the magnet. RADIA code produces the kick map from three dimensions of the analytical EPU field. The TFR cause electrons to experience horizontal $\theta_x$ and vertical $\theta_y$ angular kicks when they wiggle transversely in the EPU. The angular kicks are defined

$$\begin{aligned} \theta_x &= -\frac{\partial \Phi}{\partial x} \\ \theta_y &= -\frac{\partial \Phi}{\partial y} \end{aligned} \tag{3}$$

with a focusing potential given by

$$\Phi(x,y) = -\frac{1}{2}\left(\frac{e}{\gamma mc}\right)^2 \int_{-\infty}^{\infty}\left(\left(\int_{-\infty}^{s} B_x(x,y,s_1)ds_1\right)^2 + \left(\int_{-\infty}^{s} B_y(x,y,s_1)ds_1\right)^2\right)ds \tag{4}$$

where $e/\gamma mc$ is the beam rigidity, and $B_x$ and $B_y$ are the magnetic flux density in horizontal and vertical planes, respectively. The 3D kick map (Fig. 3) for the horizontal polarization mode of the EPU shows that the angle between the transverse trajectory and these fields causes the electron to experiences local focusing that depends on the electron positions. The focal lengths are

$$\begin{aligned} \frac{1}{F_x} &= -\frac{1}{2}\left(\frac{e}{\gamma mc}\right)^2 \int_{-\infty}^{\infty} \frac{\partial^2}{\partial x^2}\Phi(x,y,s)ds \\ \frac{1}{F_y} &= -\frac{1}{2}\left(\frac{e}{\gamma mc}\right)^2 \int_{-\infty}^{\infty} \frac{\partial^2}{\partial y^2}\Phi(x,y,s)ds \end{aligned} \tag{5}$$

These focusing properties depend on the polarization mode introduce tune shifts given by

$$\Delta Q_{x,y} = \frac{1}{4\pi}\frac{\beta_{x,y}}{F_{x,y}} \tag{6}$$

For the PLS-II, the EPU72 is installed at a long straight section with beta functions ( $\beta_x$ / $\beta_y$ = 6.6 m/4.4 m). In the three polarization modes, the tune shifts differently as a function of the horizontal axis (Fig. 4).

The magnetic fields of the EPU also perturb the betatron functions along the ring. The perturbations affect the tune and betatron function in horizontal and vertical plane in ways that depend on the polarization modes at the minimum gap (Table 4). Beta beating by EPU is not significant. Compared with EPUs in other facilities [8], EPU72 in PLS-II have much smaller horizontal and vertical angular kicks (Fig. 5) (order of one). These results imply that EPU72 will not reduce the lifetime or cause injection problems in PLS-II.

## IV. FREQUENCY MAP ANALYSIS

To track the particles, the angular kicks in both planes were generated using 3D RADIA magnet modelling code, then mapped and imported into Accelerator toolbox [9], which is based on MATLAB. 6D particle tracking and frequency map analysis (FMA) were performed to identify some harmful resonances caused by the non-linear field roll-off of the EPU72. The FMA converts the frequency map from spatial; domain of initial position to the frequency domain over a finite time by numerical integration of motion in the initial coordinate space $(x_i, y_i, p_{xi}, p_{yi})$. The phase trajectory is integrated through one turn, then stored. After a sufficient number of turns, the spatial data are converted to the frequency domain ( $\upsilon_x, \upsilon_y$ ) by a Fourier Transformation based on Numerical Analysis of the Fundamental Frequency (NAFF) [10]. With particle tracking for 1024 turns, the dynamic aperture in coordinate space and in tune space for the PLS-II with the EPU were obtained for three polarization modes (Fig. 6). The figure indicates sufficient dynamic apertures of up to 15 mm in the horizontal plane and up to 4.5 mm in the vertical aperture. Therefore, EPU72 has no significant effect on beam stability, especially its lifetime and injection.

## V. CONCLUSION

Frequency map analysis was used to quantify nonlinear effects of an Elliptical Polarization Undulator (EPU72) on beam dynamics in PLS-II. This device was designed to produce a photon beam with an energy of 1 – 5 keV. A kick map, and frequency map analysis confirm that EPU72 does not significantly degrade beam stability.

## ACKNOWLEDGEMENT

We thank H. Wiedemann for useful discussions and guidance, and W. Wan for guidance in code-tracking. This research was supported by the Converging Research Center Program through the Ministry of Science, ICT and Future Planning, Korea (NRF-2014M3C1A8048817) and the Basic Science Research Program through the National Research Foundation of Korea (NRF-2015R1D1A1A01060049).

## REREFENCES

Table 1. Main parameters for PLS-II.

| Parameter | Value | Unit |
|---|---|---|
| Beam energy | 3 | GeV |
| Beam emittance | 5.8 | nm |
| Beam current | 400 | mA |
| ID straight section | 20 | |
| Total beamlines | > 40 | |
| Injection mode | Top-up | |
| Absolute orbit control | < 100 | μm |
| Orbit stability | < 1 | μm |

Table 2. Main parameters for the EPU72.

| Parameter | Value | Unit |
|---|---|---|
| Period length | 7.2 | cm |
| Number of periods | 34 | |
| Min. magnetic gap | 18 | mm |
| Permanent material | NdFeB | |

Table 3. Magnetic field strength for each EPU72 phase. Δz: longitudinal translation of the diagonal arrays; $B_x$: horizontal magnetic field; $B_y$: vertical magnetic field.

| Polarization mode | Bx [T] | By [T] | Deflecting parameter K |
|---|---|---|---|
| Horizontal (Δz = 0 mm) | - | 0.723 | 4.861 |
| Circular (Δz = 20.87 mm) | 0.486 | 0.486 | 3.271 |
| Vertical (Δz = 36 mm) | 0.563 | - | 3.786 |

Table 4. Linear optic perturbation caused by EPU72.

| Polarization mode | $\Delta Q_x$ | $\Delta Q_x$ | rms horizontal beta-beating [%] | rms vertical beta-beating [%] |
|---|---|---|---|---|
| Horizontal ($\Delta z$ = 0 mm) | 0.0009 | 0.0015 | 0.4 | 0.7 |
| Circular ($\Delta z$ = 20.87 mm) | -0.0014 | 0.0020 | 0.6 | 0.8 |
| Vertical ($\Delta z$ = 36 mm) | -0.0014 | 0.0018 | 0.6 | 1.0 |

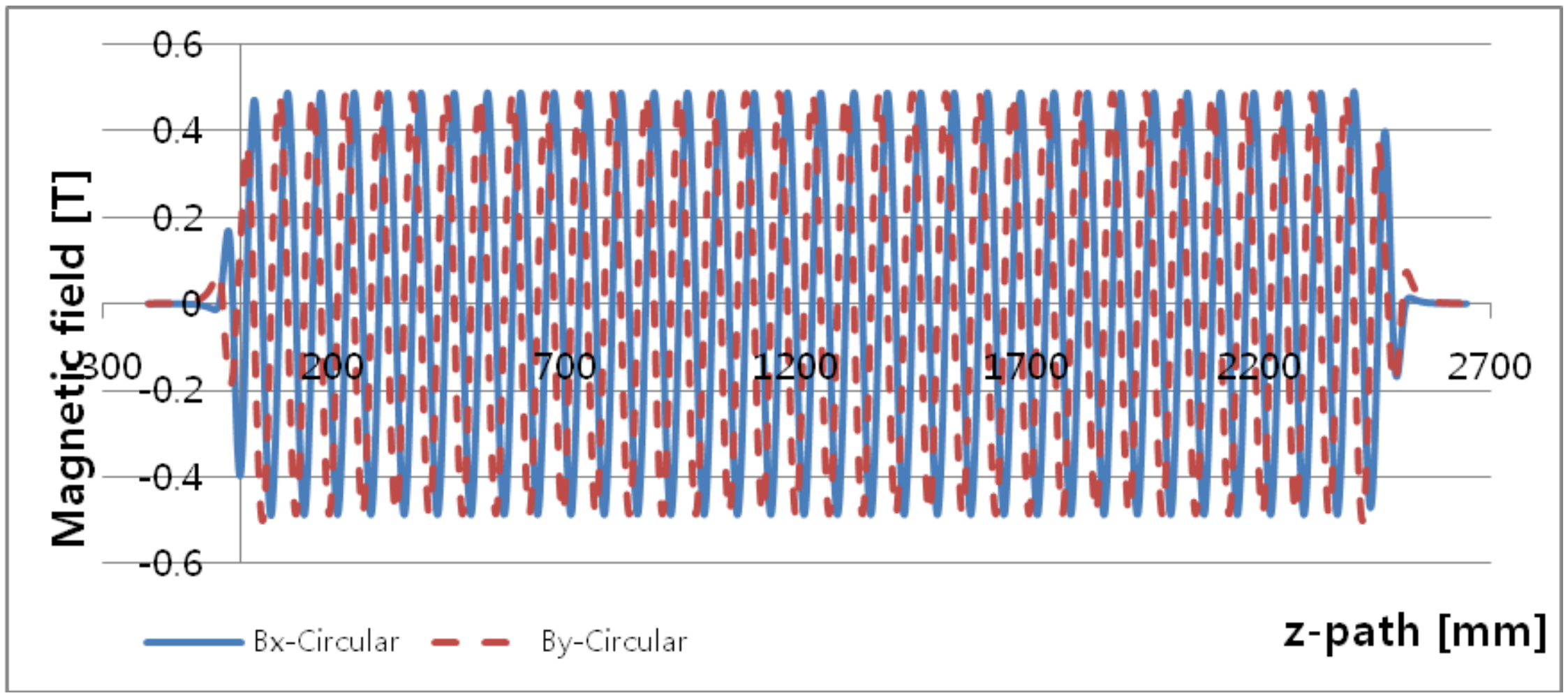

Fig. 1. Calculated magnetic field on axis in circular mode of polarization at minimum gap. Here z is longitudinal axis and Bx and By are horizontal and vertical magnetic fields, respectively.

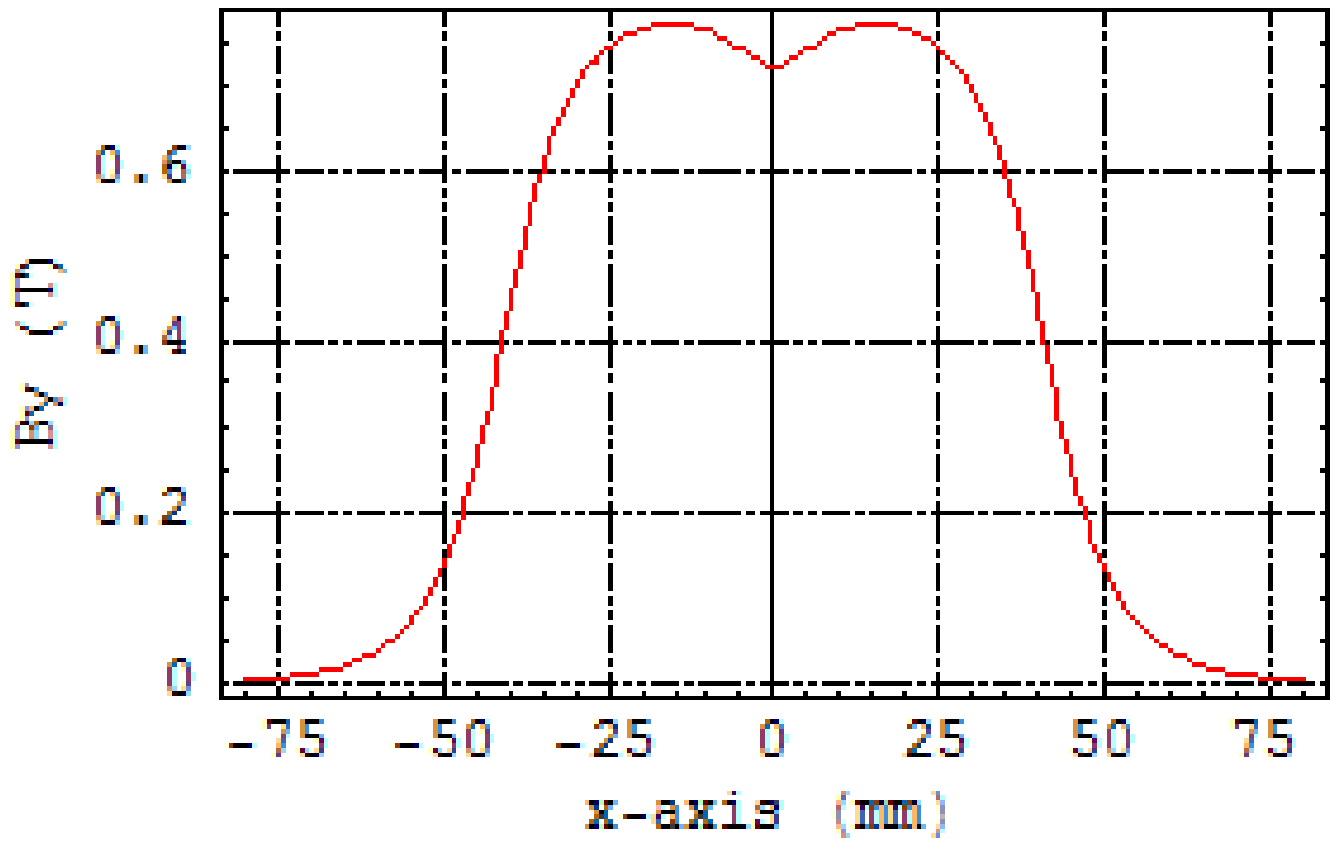

Fig. 2. Calculated vertical field for horizontal mode of polarization distribution along x-axis. Because the left and right arrays should be translated, a small (~1.0 mm) horizontal clearance is placed between left and right arrays centered at x = 0. Due to this horizontal gap, the vertical field at zero phase dips at x=0.

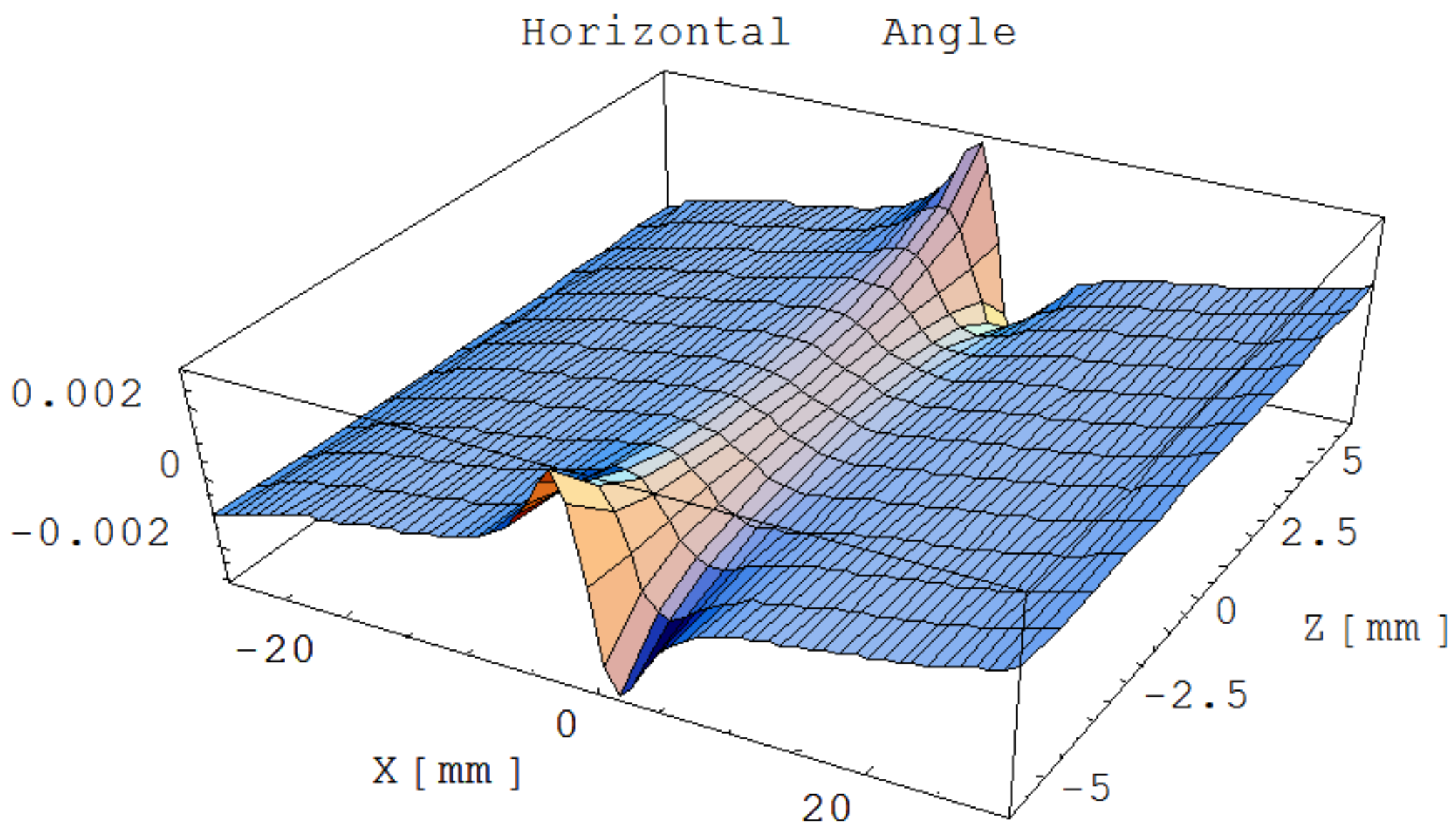

Fig. 3. 3-D horizontal kick map for EPU72 with horizontal polarization mode.

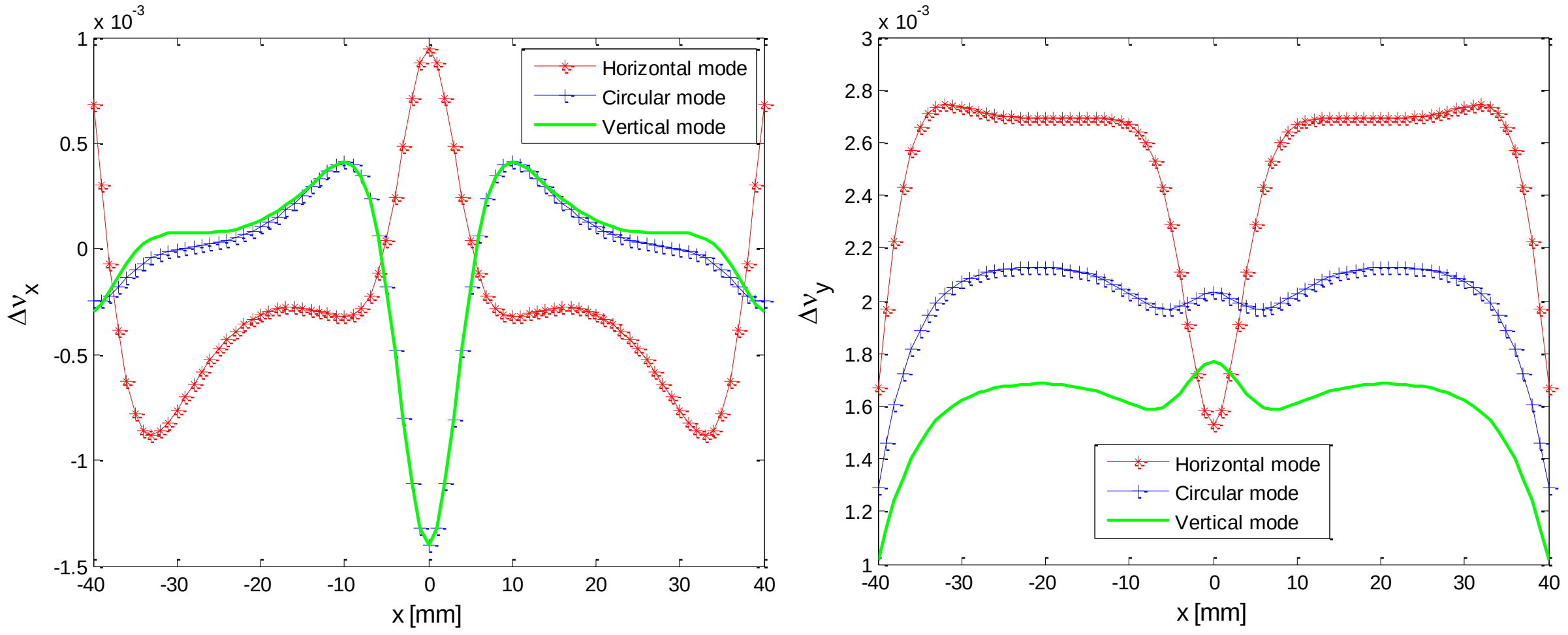

Fig. 4. Horizontal (left) and vertical (right) tune shift as function of horizontal position.

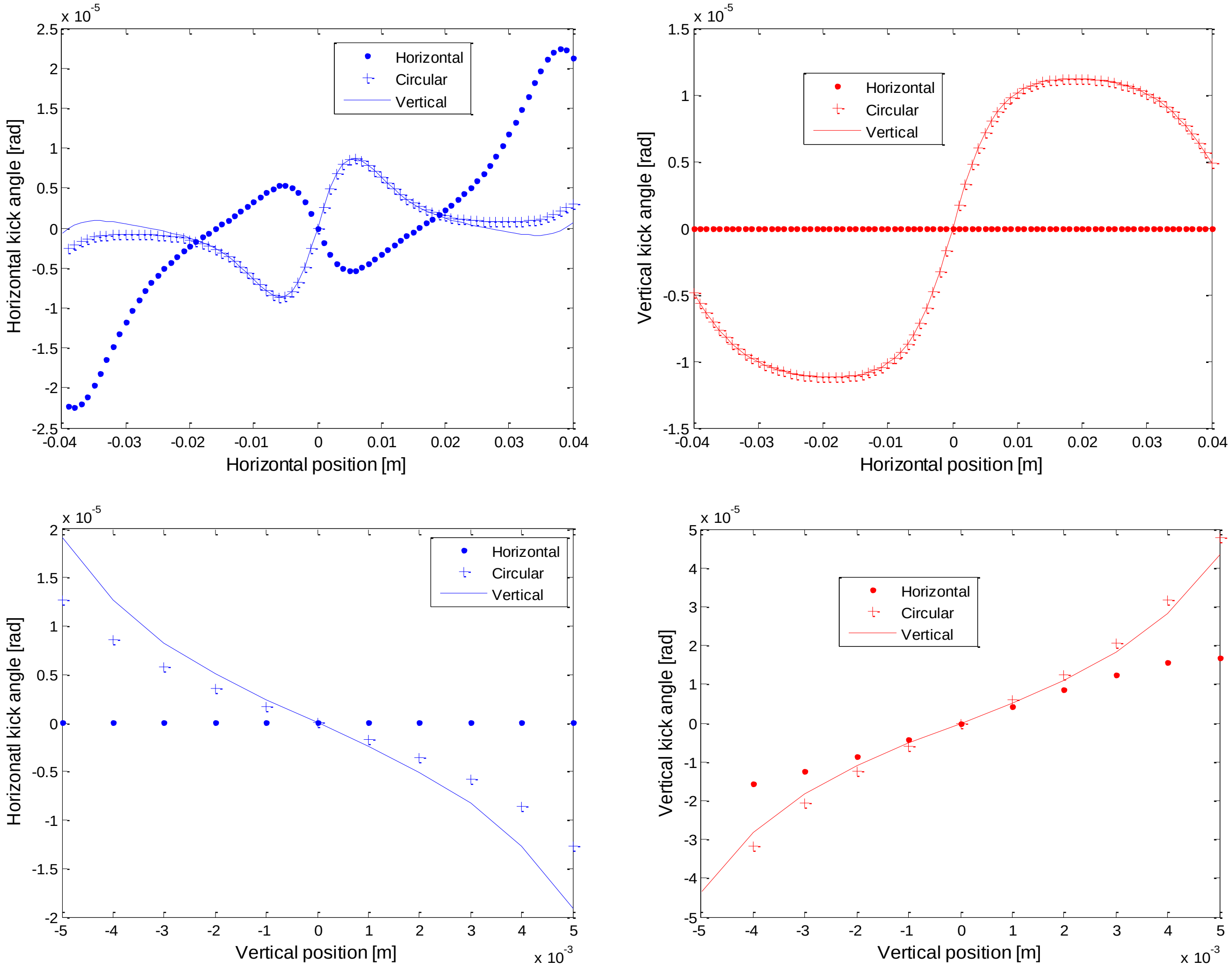

Fig. 5. Horizontal (left) and vertical (right) angular kick for horizontal (dots), circular (plus) and vertical (solid line) modes at minimum gap of 18 mm. Compared with EPUs in other facilities [8], EPU72 in PLS-II have much smaller angular kicks (order of one). This figure implies that EPU72 will not reduce the lifetime and or cause injection problems in PLS-II.

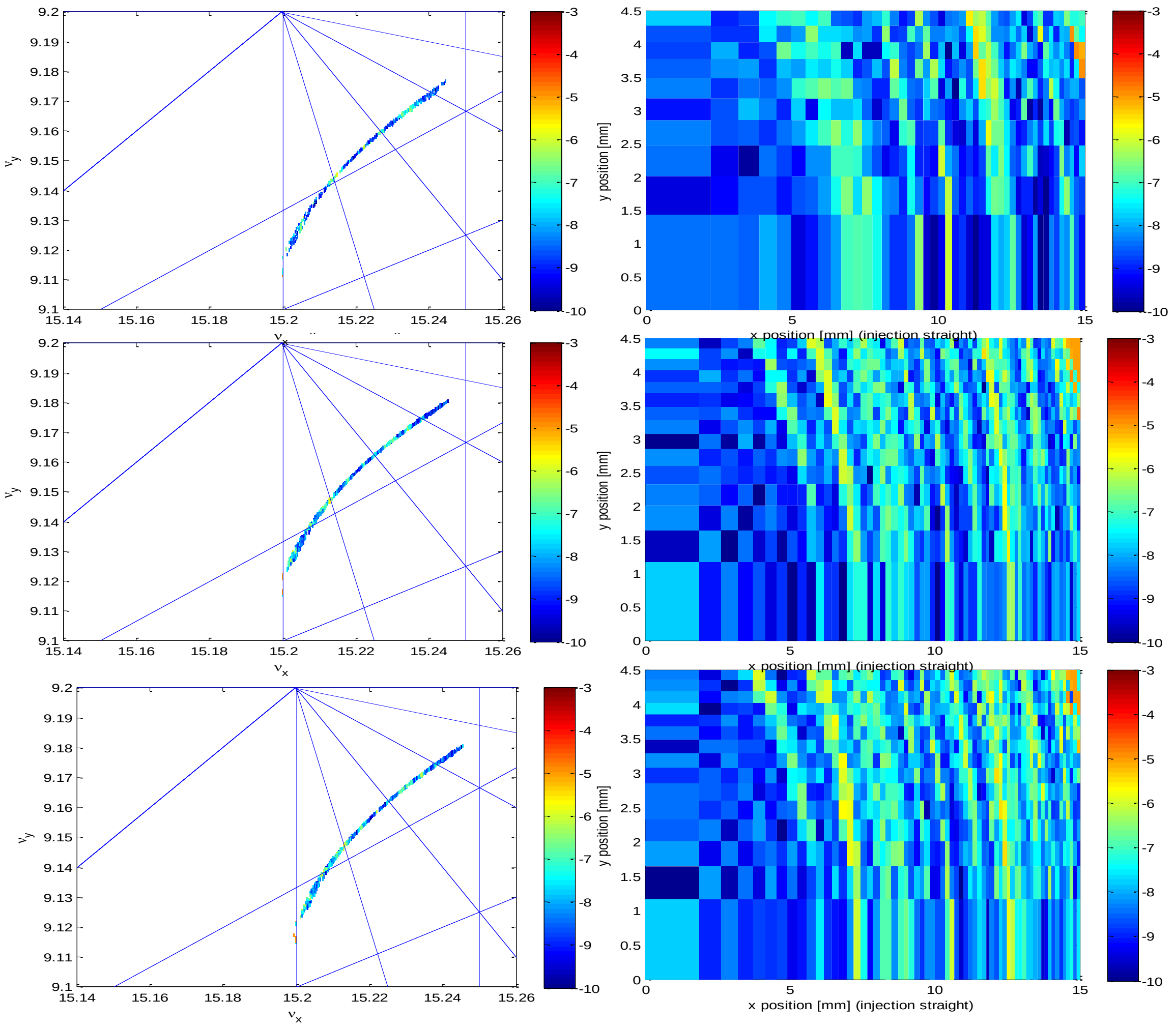

Fig. 6. Frequency map in coordinate and frequency spaces for three polarization modes with phase of 0° (top), 104.3° (middle) and 180° (bottom) at minimum gap of 18 mm. The figure indicates sufficient dynamic apertures up to 15 mm in the horizontal plane and up to 4.5 mm in the vertical aperture. Therefore, EPU72 will not significantly reduce the lifetime and or cause injection problems in PLS-II.